\documentclass[twocolumn,showpacs,amsmath,amssymb]{revtex4}
\usepackage{graphicx}
\usepackage{color}
\usepackage{graphicx,psfrag,booktabs}

\begin{document}

\title{Interlayer coupling enhanced by the interface roughness}
\author{Ching-Hao Chang and Tzay-Ming Hong}
\affiliation{Department of Physics, National Tsing Hua University, Hsinchu 30043, Taiwan, Republic of China}
\date{\today}

\begin{abstract}
Previous experiment on Fe/Cr/Fe(001) trilayers reported a surprising observation that the
interlayer exchange coupling could be enhanced drastically by the bombardment of irradiation even at low fluences.
We propose that it is due to the  resonant states in the spacer, made possible when the topography of both interfaces is correlated and exhibits prominent Fourier components.
A systematic procedure is developed to handle the interface roughness and predict on how to optimize the interlayer coupling.
This method can be extended to bridge the gap between theories and experiments in other heterojunctions.
\end{abstract}


\maketitle

\section{Introduction}
The interlayer exchange coupling has been studied for more than twenty years\cite{I1,I2,I3,MD,Kudrnovsky,Bruno} with applications in phenomena such as Giant Magnetoresistance\cite{g1,g2} (GMR) and Tunnelling Magnetoresistance (TMR). When it comes to estimate the effect of interface roughness (IR) on the coupling, there are few microscopic theories other than resorting to static average\cite{MD,JO,JU,Bruno2}.
A recent experiment\cite{bomb} reported that the strength of the coupling
in Fe/Cr/Fe(001) trilayers could be modified in a controlled manner and enhanced even at low fluences by the ion beam irradiation.
The ion bombardment is known to be an excellent tool for patterning magnetic areas on interfaces of multilayer system without changing the sample structure\cite{sc}.

This enhancement of interlayer coupling is not expected by the conventional approach to statistically average over the IR.
Many experiments\cite{bomb,pattern1,pattern2} have been dedicated to address this surprising finding. They tailored the interface into exhibiting certain patterns by the ion beam irradiation, but so far there is no conclusive result\cite{pattern1,pattern2}. 
Among the many possible coupling mechanisms in the trilayer systems that might give rise to this enhancement, the authors of Ref.\cite{bomb} have estimated the percentage of having the magnetic bridges and concluded that their contribution to the coupling was minor. The spin density wave in bulk chromium has been known\cite{sdw,fishman} to be crucial at understanding how the electronic structure and the topology of Fermi surface influence the magnetic properties of Fe/Cr multilayers. However, since the enhanced coupling was observed with a spacer of only 8 monolayers\cite{bomb}, the indirect exchange is more likely to be mediated by the conventional Ruderman-Kittel-Kasuya-Yosida interaction\cite{Bruno} instead of spin density waves which occur\cite{fishman,cr} only above a critical thickness of 30 monolayers. Another possible candidate for the coupling is magnetic dipole interactions\cite{dipole}, made possible by IR. Nevertheless, it is not likely to dominate the coupling because the analysis of STM images showed that the thickness fluctuations of the spacer in Ref.\cite{bomb} was only one monolayer.

In this article, we present the first microscopic theory based on a perturbative treatment of the IR. This modifies the wavefunction of the conduction electrons in the spacer and the interlayer coupling they mediate. Mainly, we found two additional terms to the scattering wavefunctions due to the topography of IR besides the usual static average results. How and when these extra terms can lead to enhancement of the interlayer coupling is addressed.

In section II, we discuss the energy spectrum of quantum well states (QWS) in a trilayer system, which is shown to be broadened by the rough interface, as expected\cite{Chiang} from taking the static average.
In addition, when the two interfaces are assigned a common periodic variation,
the spectrum will get shifted upwards.
In section III, a resonance is predicted to occur when
the length scale of IR matches the spacer width or the Fermi wavelength. 
We propose that the latter matching is responsible for the enhancement of interlayer coupling by the ion-beam bombardment. Conclusions and discussions are arranged in section IV, in which other coupling mechanisms, such as the magnetic dipole interaction and the biquadratic coupling, and the effect of alloying at the interface are addressed. To preserve the conciseness of the main text, detailed calculations are all arranged in the Appendices.

\section{Rough interface and corrections to quantum well states}
We start by studying a simple 2-D case: an interface described by $x= A \sin (p y)$
and assume the effective mass to be the same for carriers at different sides of the interface.
As a plane wave comes in with momentum $(k_x, k_y)$ from the left,
we assume the amplitude $A$ of the roughness to be much smaller than both $1/k_x$ and $1/p$. The $k_{x} A \ll 1$ allows us to treat the roughness as a perturbation, while $pA  \ll 1$ precludes the possible emergence of local states at the interface, as in water waves along the coastline.
Under this limit,
the action of static averaging over one rough interface can be shown to be equivalent to the scattering of two effective smooth interfaces, the first and second terms in Eq.(\ref{eq:rough}).
Whiles, the topography at the interface creates the third and fourth terms, of which the strength is
proportional to $A$ and $k_y$ is shifted by $p$:
\begin{align}
\notag
\Phi_{in}\approx &\frac{1}{2} \Big[e^{i\frac{k_{x}A}{\sqrt{2}}}\Phi_{in}^{0}(x-\frac{A}{\sqrt{2}},y)+e^{-i\frac{k_{x}A}{\sqrt{2}}}\Phi_{in}^{0}(x+\frac{A}{\sqrt{2}},y)\Big]\\
&+a^{(1)}_{\vec{q}1}e^{-iq_{1x}x+i(k_{y}-p)y}+a^{(1)}_{\vec{q}2}e^{-iq_{2x}x+i(k_{y}+p)y}
\label{eq:rough}
\end{align}
where $\Phi_{in}^{0}(x,y)$ denotes the wave function scattered by a smooth interface and
$a^{(1)}_{\vec{q}1,2}$ are defined in Appendix A which contains details of the first-order perturbation calculations.

In a trilayer system, QWS determine the interlayer exchange coupling.
It is crucial to study the variation of QWS due to the IR, which broadens the resonance peaks, shifts the energy spectrum, and sometimes reduces the number of bound states. We find that the energy shift becomes large when (1) $1/p$ is comparable to the thickness of the spacer $D$ and (2) the topographies on these two interfaces are correlated, which feature could not be obtained by the static average\cite{MD,JO,JU} or another approach which combined it with the Green's function\cite{glutsch}.
An example is shown in Fig.1, assuming both interfaces exhibit the same topography $x=A \sin(p y)$.
From Appendix C, it is shown that the only effect of static average is to broaden the peaks in the energy spectrum that represent bound states.
When the first-order perturbation corrections due to the IR are added,
 the spectrum becomes shifted and the original bound states near the barrier edge can become unbounded. This shift has been observed in the photoemission experiment\cite{Chiang}.
Note that both features are more pronounced at high energy states because the shorter wavelength of QWS  renders them more susceptible to the topography of the interface.
We shall discuss later our finding that the $p\approx 1/D$ that causes the biggest spectrum shift often brings about the greatest enhancement or suppression of the interlayer coupling.
Other details, such as the interface intermixing and Brillouin zone dilation\cite{period},
seem not to be crucial for qualitative predictions.

\begin{figure}
\includegraphics[width=0.49\textwidth]{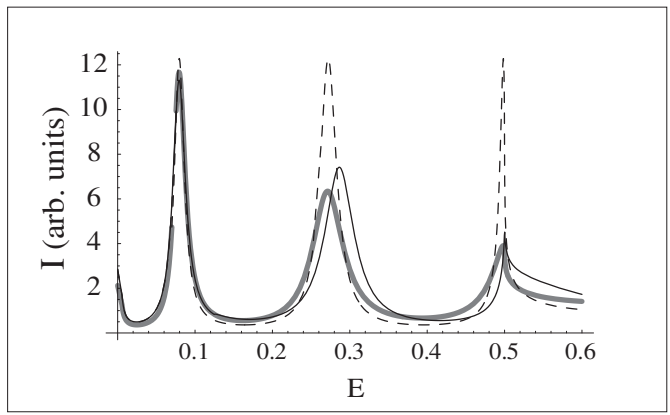}
\caption{Photoemission intensity is plotted as a function of the energy of quantum well states  when both interfaces exhibit the same topography $x=2A \sin(p y)$. The results for smooth interfaces are plotted in dashed line for comparison. The gray line takes care of the IR by static average, while the solid line further considers the corrections due to the first-order perturbation.
The parameters chosen for this plot are $m^\star =1$,
the depth of quantum well is $V_0=0.5$ in unit of $Ry/\pi^2$, $2A=0.5$ in unit of the Bohr radius, $p = 0.5$, broadening factor defined in Appendix C is set to be 1,
and $D=6.8$.}
 \label{resonance}
\end{figure}

A similar feature is found in the scattering states, namely, interlayer coupling can be enhanced or suppressed by the IR when (1) the topographies on both interface are correlated and (2) $p$ happens to be comparable to one of the wavelengths $k$ that satisfy the condition for the Fano resonance. Although the shift of these $k$ by IR is small, they somehow collaborate to creat a profound effect on the coupling.

\section{Interlayer Exchange Coupling}
In this section, we will study the effect of IR on the interlayer exchange coupling for a 3-D trilayer system, which consists of a spacer of width $D$ and potential $V=0$ sandwiched between layers L and R
with potential $V_{L}$, $V_{R}$, respectively.
Assuming the lattice structure along the planar directions of the interface is ideal, the coupling strength is a superposition of the contributions from all the QWS in the interface Brillouin zone\cite{Bruno,MD} (IBZ). The existence of interfaces without the IR affects the total electron energies of the spacer, the shift of which equals
\begin{equation}
\triangle E
=\frac{ -1}{4\pi^{3}}{\rm Im}\sum^\infty_{n=1}\frac{1}{n} \int_{-\infty}^{E_{F}}dE\int_{IBZ}d^{2}\vec{k}_{\|}
r_{L,k_{\perp}}^{n}r_{R,k_{\perp}}^{n}e^{2ink_{\perp}D}
\label{eq:interlayerE}
\end{equation}
 per unit area of interface. The reflective coefficients $r_{L/R}$ at the left/right interface are functions of the wave number along the normal direction of the interface $k_{\perp}$, which depends on the energy $E$ and $\vec{k}_{\|}\equiv \vec k-\vec k_{\perp}$. The integer $n$ denotes the number of round-trip reflections by the two interfaces.
In the presence of IR, two corrections are introduced to Eq.(\ref{eq:interlayerE}):
static average terms in Eq.(\ref{eq:rough}) come in through the $r_{L/R}$, while the last two terms due to the perturbation contribute additional close loops when the topography of both interfaces are correlated. By use of the T-matrix and perturbation methods, we derive the energy shift in Appendix B, which takes the form:
\begin{align}
\notag &\triangle E^{r}\approx  \frac{2}{(2\pi)^{3}}{\rm Im} \int_{-\infty}^{E_{F}}dE\int_{IBZ} d^{2}\vec{k}_{\|}\\
\notag  &\Big[\sum_{n}\frac{-1}{n}r_{R,k_{\perp}}^{n}(1-d_{L}^{2}k_{\perp}^{2})^{n}r_{L,k_{\perp}}^{n}(1-d_{R}^{2}k_{\perp}^{2})^{n}e^{2ink_{\perp}D}\\
& +\sum_{q_{y}} b^{(1)}_{L,k_{y}q_{y}}b^{(1)}_{R,q_{y}k_{y}}e^{i(q_{\perp}+k_{\perp})D} \Big]
\end{align}
where $d_{L/R}$ are corrections after averaging the IR on the left/right interface, and $b^{(1)}_{L,k_{y}q_{y}}$ ($b^{(1)}_{R,q_{y}k_{y}}$) are additional transmission coefficients from momentum
$\{k_{\perp},k_{y}\}$ to $\{q_{\perp},q_{y}\}$ (vice versa) induced by the tomography on interface $L$ ($R$). We checked that the above equation indeed reduced to Eq.(\ref{eq:interlayerE}) at the limit of $p\rightarrow 0$.

An important application of trilayers is the GMR and TMR systems where the two side layers exhibit permanent magnetization.
When they are ferromagnetically coupled, the energy shift by Eq.(\ref{eq:interlayerE}) can be calculated by evaluating the reflective coefficients under the corresponding Zeeman field. It is found to differ from that of the antiferromagnetic case.
The magnetic coupling strength $J(D)$ can then be defined as the difference between these two energies\cite{Bruno}.
For simplicity, we assume the two magnetic layers to be of the same material and their interface topographies are correlated. After some routine calculations in Appendix B, the magnetic coupling is found to be:
\begin{align}
\notag& J(D)\approx \frac{2}{(2\pi)^{3}}{\rm Im} \int_{-\infty}^{E_{F}}dE\int_{IBZ} d^{2}\vec{k}_{\|}\\
\notag&\Big\{\sum_{n}\frac{-1}{n}\big[r_{p}^{n}(1-d^{2}k_{\perp}^{2})^{n}-r_{ap}^{n}(1-d^{2}k_{\perp}^{2})^{n}\big]^{2}e^{2ink_{\perp}D}\\
&+ \sum_{q_{y}} \big[b^{(1)}_{p,k_y q_y}-b^{(1)}_{ap,k_y q_y}\big]
\big[b^{(1)}_{p,q_y k_y}-b^{(1)}_{ap,q_y k_y}\big]e^{i(q_{\perp}+k_{\perp})D}\Big\}
\label{eq:corret-magnetic}\end{align}
where $r_{p}$/$r_{ap}$ denote the reflective coefficients for carriers with spin parallel/antiparallel to the magnetization of the layer they reflect from.

\subsection{Uncorrelated interfaces}
An intuitive equation, which resorts to static average over the IR, has been widely used by experimentalists and theoretists alike\cite{MD,JO,JU,Bruno2}:
\begin{equation}
\bar{J}(D)=\sum_{n}w(n)J(nd)
\label{staticaverage}\end{equation}
where $w(n)$ is the weighting of $n$-layers among the various widths bewteen two rough interfaces,  $d$ is the width of one unit layer in the spacer, and $D$ is defined as the mean width $\sum_n w(n) nd$. This weighting $w(n)$ is
usually assumed to obey the Guassian distribution\cite{MD,JO,JU,Bruno2}.

\begin{figure}
\includegraphics[width=0.5\textwidth]{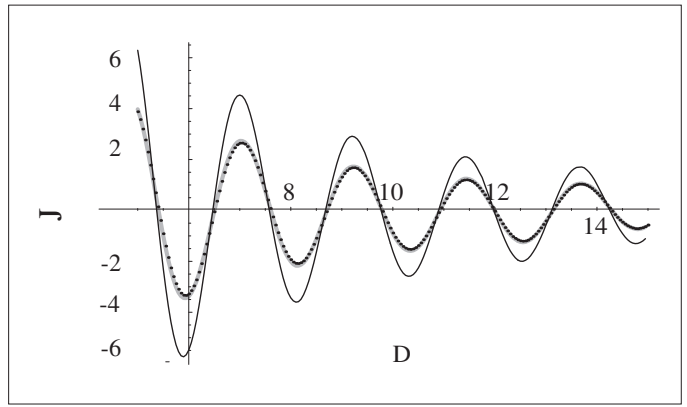}
\caption{Coupling strength in unit $10^{-5}{Ry}/{\pi^{2}}a_{0}$ between rough interfaces with random topography in 2-D is calculated by Eq.(\ref{eq:corret-magnetic}) (in dashed line) and the static average Eq.(\ref{staticaverage}) (in gray line) as a function of  the average width $D$ in Bohr radius $a_{0}$. The parameters we used are: the Fermi energy $E_{F}=1$, $V_{p}=0.01$ and $V_{ap}=0.1$ (energy unit is $Ry/\pi^{2}$), and the maximum amplitude of the IR on both interface is $a_{0}/3$. For comparison, the result for smooth interfaces is plotted in solid line.}
\label{rough-pre}
\end{figure}

When the IR on the two interfaces are uncorrelated, the last term in Eq.(\ref{eq:corret-magnetic}) due to their interference vanishes.
The remaining term is shown in Fig.\ref{rough-pre} to be equivalent to the static average.
It can be seen from the figure that uncorrelated IR always reduces the coupling strength, but retains the same period as the exact result for smooth interfaces, which turns out to be roughly $\pi/k_F$ as predicted by the Ruderman-Kittel-Kasuya-Yosida formula - a second-order result.

\subsection{Correlated interfaces}
\begin{figure}
\includegraphics[width=0.47\textwidth]{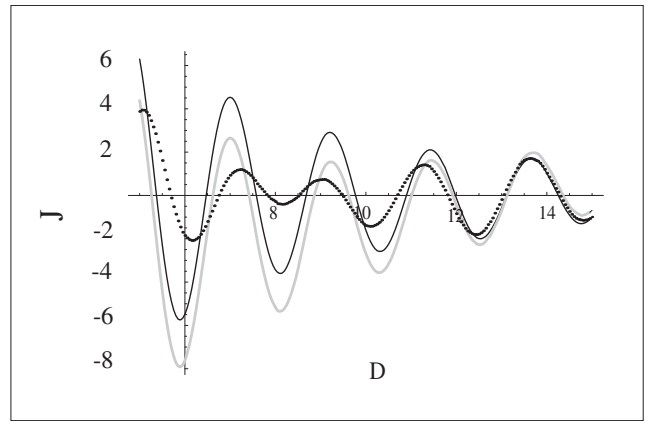}
\caption{Coupling strength is plotted as a function of the average film width $D$ as Fig.\ref{rough-pre} with the same parameters except that the two interfaces are now assumed to share the same topography $A(y)=1/3\sin(p y)$. Again, the result for smooth interfaces is shown in solid line for comparison. The gray/dashed lines represent $p=0.05/1.5$, respectively.}
 \label{couple-D}
\end{figure}

When the topography of both interfaces is correlated, the close loops in Eq.(\ref{eq:corret-magnetic}) formed by the interference will survive and give rise to new signatures.
The correlation can be introduced\cite{period} by the ion irradiation through the trilayer, which has the capacity of rearranging the atoms on both interfaces.
Although Ref.\cite{bomb} used an ionized helium of 5 keV, smaller than the  2-MeV Au$^{+2}$ in Ref.\cite{period}, we believe this correlation is still possible because the penetration depth is inversely proportional to the cross section of the incoming ions.
Consider the extreme case that both interfaces share the same  topography,  $A(y)=\sin(p y)/3$ in unit of $a_{0}$. The coupling strength is plotted in Fig.\ref{couple-D} with the same parameters as Fig.\ref{rough-pre}.

Since the gray and dashed lines are for different $p$, it is not surprising that they oscillate with different period. However, when $D$ is much larger than the amplitude of both topography, the roughness becomes immaterial and both lines start to merge with the solid line for smooth interfaces. This is consistent with the experimental observation that the effect of irradiation diminishes when the spacer gets thicker\cite{bomb}.
The important feature to note here is that, unlike Fig.\ref{rough-pre} where uncorrelated roughness always diminishes the coupling strength, the coupling here can be enhanced at certain $D$, as was observed experimentally\cite{bomb}.

\begin{figure}
\includegraphics[width=0.47\textwidth]{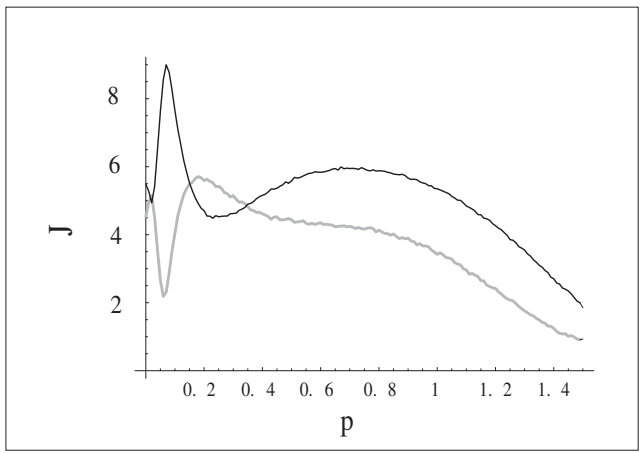}
\caption{Absolute value of the coupling strength for the same 2-D trilayer in Fig.\ref{couple-D} is replotted as a function of $p$ for a fixed $D$. The solid/gray lines are for $D=6/7\ a_{0}$, respectively.
The resonance peaks appear near $p\approx 1/D$.}
 \label{couple-resonance}
\end{figure}

The coupling constant $J$ is plotted in Fig.{\ref{couple-resonance}} as a function of $p$ for two fixed values of $D$.
Two signatures are worth noticing: (1) both lines exhibit two peaks - the sharp one has been discussed in section II to occur at $p \approx 1/D$ and is caused by the bound state, while the broad one at $p\approx k_F$ comes from the scattering state where we chose $k_{F}\approx 1$;
(2) the coupling strength decreases when $p> 1$.

\begin{figure}
\includegraphics[width=0.47\textwidth]{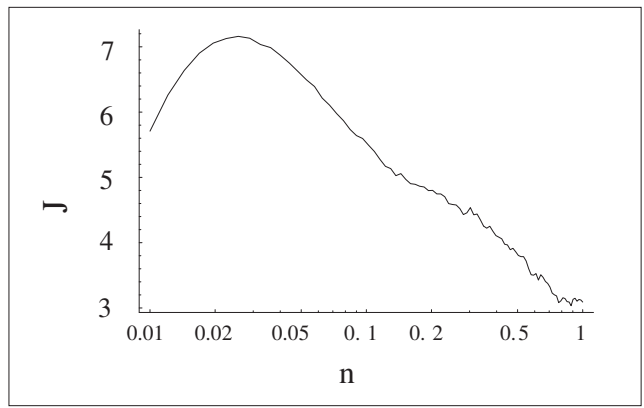}
\caption{Absolute value of the coupling strength (in unit of ${10^{-5}Ry}/({2 \pi^{3}a_0^2})$) is plotted as a function of ion dosage (in unit of ions/$a_0^2$) for a 3-D trilayer.
Same parameters as the last three figures. The average spacer width is $D=7 a_{0}$
and the topography is $A(y,z)=1/3 \sin(\sqrt{n}\ y)\sin(\sqrt{n}\ z)$.
The resonance peak appears near ``$n=0.0025$", consistent with the condition for the bound state to become resonant, and so is hard to discern in the figure. This peak is sharper than that of Fig.{\ref{couple-resonance}} for a 2-D case with a magnitude close to zero.}
 \label{3D}
\end{figure}

Let us now generalize to a 3-D trilayer. Again, assume the two interfaces to share the same topography, $A(y,z)=1/3 \sin(\sqrt{n}\ y)\sin(\sqrt{n}\ z)$, after being irradiated by ion beams with fluence $n$. The variation of the coupling strength with respect to $n$ is plotted in Fig.\ref{3D}, which resembles the gray line for the 2-D case in Fig.\ref{couple-resonance}. The position of the first peak is shifted down to such a small value that it is hard to discern it in the figure.
Note that both the qualitative behavior and position of the second peak in Fig.\ref{3D} agree excellently with the solid-circle line in Fig.3 of Ref.\cite{bomb} by Demokritov {\it et al.}
The reason why we did not extend the ion dosage to higher values is that they violate the requirement $pA\ll 1$ mentioned in the beginning of section II for the perturbation method to work.

\section{Conclusions and discussions}
In conclusion, we have developed a perturbative method to handle the interface roughness.
The broadening of the peaks in the spectrum and the usual trend of suppressing the interlayer coupling can be calculated quantitatively by our procedures, but they are already captured by the previous method of static average.
What is special about our method is that it can predict an enhancement in the interlayer coupling, as has been reported expermentally,
when the Fourier transformation of the topography on both interfaces is dominated by the same prominent component $p$ which is comparable to $k_F$. We ascribe it to the scattering states under such a correlation. In the mean time, if $p$ is close to the inverse of  the average spacer width, we show that the bound state, namely, the peaks in the spectrum will be shifted and give rise to another sharper anomaly in the interlayer coupling, which can be either an enhancement or suppression depending on the specific value of $D$ and $p$. Under the classical mechanism of dipole interactions\cite{dipole}, correlations between the topography on both interfaces have been studied to give rise to an enhanced coupling. However, it is precipitated by the largeness of IR amplitude instead of when $p\approx 1/D$ or $k_F$.

Biquadratic coupling\cite{biqi,MD} has been known to coexist in the trilayers with the bilinear one that we calculated in this work. For completeness, we have also checked the effect of IR on the  intrinsic biquadratic coupling by the perturbative approach. It is found to be enhanced as well under the same conditions as the bilinear term. However, the enhanced strength remains three orders of magnitude smaller than the latter, which conclusion is similar to that reported\cite{Bruno} for a smooth interface and without the intralayer coupling.

Another common feature linked to the ion-beam irradiation is the alloying effect. Although this is more likely to take place at the interface, the possibility of its happening in the whole spacer can not be excluded. When this occurs at some special composition, the nesting effect\cite{nesting} might come about and give rise to an enhanced interlayer coupling. The effect of alloying at the interface is considered in Appendix D. It is found to always suppresse the interlayer coupling, consistent with the conclusion of {\it ab initio} calculations\cite{alloy}. When the interdiffusion is severe, the reduction is exponential. If the extent of alloying is narrow, the correction is small and the relative enhancement due to the correlations is the same as in the absence of the alloying.

The excellent agreement of our conclusions and figures did not rely on any fitting parameter, and the microscopic topography of the interface roughness is our only input.
We thus conclude that the enhancement of interlayer coupling strength, the prime interest of this work, does not depend critically on the anisotropic field due to the interface roughness, nor the possible new channel for direct exchange between the magnetic side layers which is made possible by the irradiation\cite{bomb}.

We thank Professors John Cheng-Chung Chi and Raynien Kwo and Mr. C. H. Chen for useful comments, and acknowledge the support by National Science Council in Taiwan under grants 95-2120-M007-008 and 96-2120-M-007-002.

\begin{appendix}
\section*{Appendix}
\section{Perturbative method of scattering}
In order to realize the IR in a microcsopic way, we start from the microscopic Hamiltonian for a 2-D heterojunction with an irregular topography, $A(y)$, and the
potential on the right side is set to be higher by $V_{0}$. 
The wavefunctions on the left/right sides are respectively denoted by
$\Phi(x,y)$ and $\Psi(x,y)$. The boundary conditions are:
\begin{align}
\notag
&\Phi(A(y),y) = \Psi(A(y),y)  \\
&\frac{\partial\Phi(x,y)}{\partial x}\Big|_{x=A(y)}=\frac{\partial\Psi(x,y)}{\partial x}\Big|_{x=A(y)}
\label{a1}
\end{align}
In order to find the scattering wavefunctions, let us consider a plane wave with
momentum $(k_x, k_y)$ moving from the left towards the
interface. In the limit that $A\ll 1/k_x, 1/p$, the scattering states can be obtained by treating the IR as a perturbation to the
smooth interface:
\begin{align}
\notag
& \Phi(x,y)=\Phi_{0}(x,y)+\sum_{q_y}a_{k_{y},q_{y}}e^{-iq_{x}x+iq_{y}y}\\
& \Psi(x,y)=\Psi_{0}(x,y)+\sum_{q_y}b_{k_{y},q_{y}}e^{iq_{x}^{'}x+iq_{y}y}
 \label{p:wave}
\end{align}
where $\Phi_{0}(x,y)$ and $\Psi_{0}(x,y)$ are the original scattering states for a
smooth interface, and $\Psi_{0}(x,y)$ carries momentum $(k^\prime_x, k_y )$. For an elastic scattering, the dispersion relation in Eq.(\ref{p:wave}) is:
\begin{align}
\notag E=&\frac{k_x^2+k_y^2}{2m^{*}}=\frac{k^{\prime 2}_x+k_y^2}{2m^{*}}+V_0\\
=&\frac{q_x^2+q_y^2}{2m^\star}=\frac{q_x^{\prime 2}+q_y^2}{2m^\star}+V_{0}
\label{p:dispersion}
\end{align}
where $m^\star$ denotes the effective mass of the charge carriers. Insert Eq.(\ref{p:dispersion}) into Eq.(\ref{p:wave}), and use
$k_{x}A(y)$, $k'_{x}A(y)$ as the perturbation to expand
the boundary conditions in Eq.(\ref{a1}). We shall retain only up to the second order in perturbation for $a_{k_{y},q_{y}}$ and $b_{k_{y},q_{y}}$:
\begin{align}
\notag a_{k_{y},q_{y}}^{(1)}&=-i(q'_{x}-q_{x})T_{k_x,k_x^\prime}\mathcal{F}[A(y),q_{y}-k_{y}]\\
&=b_{k_{y},q_{y}}^{(1)}\label{p:a1a21}\\
a^{(2)}_{k_{y},q_{y}}&=\frac{1}{q_{x}+q^{'}_{x}}[(k^{'}_{x}+q^{'}_{x})\alpha+\beta]\label{p:a1a22}\\
b^{(2)}_{k_y,q_y}&=a^{(2)}_{k_{y},q_{y}}-\alpha \label{p:a1a23}
\end{align}
\begin{align}
&\alpha =T_{k_x,k_x^\prime}m^{*}V_{0}\mathcal{F}[A^{2}(y),q_{y}-k_{y}]\label{p:a1a24}\\
&\beta =-i 2m^{*}V_{0}\mathcal{F}[\sum_{q_{y}^\prime}a_{k_{y},q_{y}^\prime}^{(1)}e^{iq_y^\prime y},q_{y}
]
\label{p:a1a25}
\end{align}
where the superscripts $(1/2)$ denotes the first/second order result, the subscript ``$k_{y},q_{y}$" denotes scattering from $k_y$ to $q_y$ state, $T_{k_x,k_x^\prime}$ is the transmission coefficient for a smooth interface,
and ``$\mathcal{F}\big[A(y),q_{y}-k_{y}\big]$" represents the
Fourier Transformation of function $A(y)$ to the momentum
space in $q_{y}-k_{y}$ .

When $\mathcal{F}[A(y),p]$ contains many $p$ components, their contributions to the third and fourth terms in Eq.(\ref{eq:rough}) tend to cancel each other at the level of first order perturbation, unless  one or two $p$ dominants. When the cancellation happens, we are forced to go up to the second order. Except when the scattered wave happens to exhibit the same momentum as the incoming wave, the different components again cancel each other and we are left with the same result as that of the static average:
For example, $A(y)=\sum_{n}
\big[B_{n}\sin(p_{n})+C_{n}\cos(p_{n}y)\big]$ contains many modes, the perturbative wavefunctions in
Eq.(\ref{p:wave}) are reduced to:
\begin{align}
\notag
\Phi(x,y) &\approx e^{ik_{x}x+ik_{y}y}+R_{k_x,k_x^\prime}(1-k^{2}_{x}d_{0}^{2})e^{-ik_{x}x+ik_{y}y}\\
\Psi(x,y) &\approx T_{k_x,k_x^\prime}\Big[1-\frac{(k_{x}-k'_{x})^{2}d_{0}^{2}}{4}\Big]e^{ik^{'}_{x}x+ik_{y}y}
\label{p:ave}
\end{align}
where $k^2_x d_{0}^{2} =k^2_x\sum_{n}\big(B_{n}^{2}+C_{n}^{2}\big)$ signifies that they originate from the second-order perturbation. The above results can be shown to be equivalent to the effect of successive scattering from two smooth interfaces with $\sqrt{2}d_0$ distance apart.
\begin{align}
\notag
&\Phi(x,y) \approx \frac{1}{2} \big[e^{i\frac{k_{x}d_{0}}{\sqrt{2}}}\Phi_{0}(x-\frac{d_{0}}{\sqrt{2}},y)+e^{-i\frac{k_{x}d_{0}}{\sqrt{2}}}\Phi_{0}(x+\frac{d_{0}}{\sqrt{2}},y)\big]\\
&\Psi(x,y) \approx \frac{1}{2} \big[e^{i\frac{k_{x}d_{0}}{\sqrt{2}}}\Psi_{0}(x-\frac{d_{0}}{\sqrt{2}},y)+e^{-i\frac{k_{x}d_{0}}{\sqrt{2}}}\Psi_{0}(x+\frac{d_{0}}{\sqrt{2}},y)\big]
\label{p:superposition}\end{align}

\begin{figure}[h!]
\hspace{10mm}\includegraphics[width=0.3\textwidth]{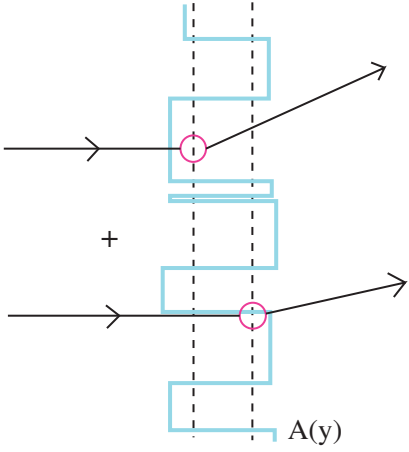}
\caption{As a plane wave is scattered by a rough interface,
the second-order effect on the wavefunction can be shown to be equivalent to the successive scattering from two smooth interfaces. }
\label{fig:ave}
\end{figure}

\section{Correction to the interlayer coupling }
Interlayer exchange coupling in Eq.(\ref{eq:interlayerE}) can be written in a more general way by the T Matrix\cite{Bruno}:
\begin{align}
\notag
\triangle E&=\frac{1}{4\pi^{3}}{\rm Im}\int_{E<E_{F}}dE\\
&\times {\rm Tr} \big[{\rm ln}(I-\hat{R}_{L}^{-+}e^{i\hat{K}^{+}D}\hat{R}_{R}^{+-}e^{i\hat{K}^{-}D})\big]
\label{p:couplingmatrix}
\end{align}
where $I$ is the unit matrix,
$\hat{R}_{L}^{-+}$/$\hat{R}_{R}^{-+}$ are the reflective matrices from the left/right smooth interfaces which is a function of energy $E$ and are nonzero only in the diagonal components, and the  superscript
$-+$ denotes that the left-going wave is turned into right-going and vice versa for $+-$.
Matrix $\hat{R}_{L}^{-+}$ takes the form\cite{Bruno}:
\begin{align}
\notag
&\hat{R}_{L}^{-+} = \left[
\begin{array}{cccc}
r_{L,k_{\perp}} & 0 & 0 & \cdots\\
0 &\ddots& & \\
0 &  & r_{L,q_{\perp}} &  \\
\vdots &  && \ddots
\end{array} \right]  \\
\label{p:energymatrix}
\end{align}
where $r_{L,k_{\perp}}$/$r_{L,q_{\perp}}$ are the reflective coefficients of momentums $k_{\perp}$/$q_{\perp}$
from the left interface. Similar expression by replacing the subscript $L$ by $R$ is the definition of $\hat{R}_{R}^{+-}$. The notation $\hat{K}^{\pm}$ denotes
\begin{align}
\notag
&\hat{K}^{+} =\hat{K}^{-}= \left[
\begin{array}{cccc}
k_{\perp} & 0 & 0 & \cdots\\
0 &\ddots& & \\
0 &  & q_{\perp} &  \\
\vdots &  && \ddots
\end{array} \right].  \\
\label{p:kmatrix}
\end{align}

When the topography $A_{L}(y)$/$A_{R}(y)$ at the left/right interfaces is considered,  the second-order perturbation result in Appendix A will introduce corrections to the reflective matrices ($\hat{R}_{L}^{-+}+\hat{\varepsilon}_{L}$)/($\hat{R}_{R}^{+-}+\hat{\varepsilon}_{R}$):
\begin{align}
\notag\hat{\varepsilon}_{L}^{-+} =& \left[
\begin{array}{cccc}
0 & \cdots & b_{L,q_{y}k_{y}} & \cdots\\
\vdots &\ddots& & \\
b_{L,k_{y}q_y} &  & 0 &  \\
 \vdots &  && \ddots
 \end{array} \right]  \\
-&\left[ \begin{array}{cccc}
r_{L,k_{\perp}}d^{2}_{L}k^{2}_{\perp} & 0 & 0 & \cdots\\
0 &\ddots& & \\
0 &  & r_{L,q_{\perp}}d^{2}_{L}q^{2}_{\perp} &  \\
 \vdots &  && \ddots
\end{array} \right]
 \label{p:per-matrix}\end{align}
and similar expression for $\hat{\varepsilon}_{R}$ after replacing the $L$ by $R$. By use of Appendix A, the weighting $b^{(1)}$ and $d$ can be copied as:
\begin{align}
\notag
&b^{(1)}_{L,k_{y}q_{y}}=i(q_{\perp}-q'_{L\perp})T_{L,\vec{k}\vec{k'}}\mathcal{F}[A_{L}(y),q_{y}-k_{y}]\\
\notag&b^{(1)}_{R,q_{y}k_{y}}=i(k_{\perp}-k'_{R\perp})T_{R,\vec{q}\vec{q'}}\mathcal{F}[A_{R}(y),k_{y}-q_{y}]\\
\notag&d_{L}= \sum_{n}\big(B_{Ln}^{2}+C_{Ln}^{2}\big)  \\
&d_{R}= \sum_{n}\big(B_{Rn}^{2}+C_{Rn}^{2}\big)
\label{p:babb}\end{align}

Insert the full reflective matrices with IR into Eq.(\ref{p:couplingmatrix}) and expand the trace term to the second order in
$\hat{\varepsilon}_{L}$/$\hat{\varepsilon}_{R}$:
\begin{align}
\notag &\triangle E^{r}\approx \frac{1}{4\pi^{3}}{\rm Im}\int_{E<E_{F}}dE\\
&\times {\rm Tr}\Big[ {\rm ln}(I-\hat{M})+\hat{\varepsilon}\frac{1}{I-\hat{M}}
-\frac{1}{2}\hat{\varepsilon}\frac{1}{I-\hat{M}}\hat{\varepsilon}\frac{1}{I-\hat{M}}\Big]
\label{p:corec-matrice}
\end{align}
where
\begin{align}
\hat{M}\equiv\hat{R}_{L}^{-+}e^{i\hat{K}^{+}D}\hat{R}_{R}^{+-}e^{i\hat{K}^{-}D}
\end{align}
and
\begin{align}
\notag\hat{\varepsilon}\equiv&\hat{\varepsilon}_{L}e^{i\hat{K}^{+}D}\hat{R}_{R}^{+-}e^{i\hat{K}^{-}D}                                      +\hat{R}_{L}^{-+}e^{i\hat{K}^{+}D}\hat{\varepsilon}_{R}e^{i\hat{K}^{-}D}\\                                                      &+\hat{\varepsilon}_{L}e^{i\hat{K}^{+}D}\hat{\varepsilon}_{R}e^{i\hat{K}^{-}D}
\label{p:per-matrix2}
\end{align}
Substituting the above two definitions into Eq.(\ref{p:corec-matrice}), the interlayer exchange coupling can be obtained as
\begin{align}
\notag\triangle E^{r}\approx & \frac{2}{(2\pi)^{3}}{\rm Im} \int_{-\infty}^{E_{F}}dE\int_{IBZ} d^{2}\vec{k}_{\|}\\
\notag & \Big[\sum_{n}\frac{-1}{n}r_{R,k_{\perp}}^{n}(1-d_{L}^{2}k_{\perp}^{2})^{n}r_{L,k_{\perp}}^{n}(1-d_{R}^{2}k_{\perp}^{2})^{n}e^{2ink_{\perp}D}\\
&+ \sum_{q_{y}} b^{(1)}_{L,k_{y}q_{y}}b^{(1)}_{R,q_{y}k_{y}}e^{i[q_{\perp}+k_{\perp}]D} \Big].
\label{P:per-result}
\end{align}

\section{Energy spectrum in rough thin films}
The photoemission intensity $I(E)$ can be used to determine the energy spectrum\cite{Chiang} of a smooth thin film, which
is a function of the reflective coefficients $r_{L/R}$ on  the left/right interfaces and the film width $D$:
\begin{equation}
I(E)\propto \left|\frac{1}{1-r_{L,k_{\perp}}r_{R,k_{\perp}}e^{i 2k_{\perp}D}+i\Lambda}\right|^{2}
\label{smoothC}
\end{equation}
where $\Lambda $ is the broadening factor\cite{Chiang} which depends on the film thickness and the mean free path. Bound states satisfy $r_{L,k_{\perp}}r_{R,k_{\perp}}e^{i 2k_{\perp}D}=1$ and correspond to
the peaks in $I(E)$.
When IR are considered, more close-loops from successfive reflections and the action of static averaging will modify both $r_{L/R}$. Eq.(\ref{smoothC}) becomes
\begin{align}
\notag I(E) \propto &\big|1-r_{L,k_{\perp}}r_{R,k_{\perp}}(1-d_{L}^{2}k_{\perp}^{2})(1-d_{R}^{2}k_{\perp}^{2})e^{i 2k_{\perp}D} \\
&+\sum_{q_{y}} b^{(1)}_{L,k_{y}q_{y}}b^{(1)}_{R,q_{y}k_{y}}e^{i[q_{\perp}+k_{\perp}]D}+i\Lambda\big|^{-2}
\label{roughC}
\end{align}
where $d_{L,R}$, $b^{(1)}_{L,k_{y}q_{y}}$, and $b^{(1)}_{R,q_{y}k_{y}}$ are defined in Eq.(\ref{p:babb}). It is checked that Eq.(\ref{roughC}) reduces to Eq.(\ref{smoothC}) when we take the limit
$A_{L,R}(y)\rightarrow 0$ or $p\rightarrow 0$.

\section{Alloying}
Besides the topography that we have discussed in detail, the reallignment of atoms at the interface due to ion beam irradiation should also include
alloying. We estimate its effect on the coupling by softening the sharp boundary at the interface into a smooth  Eckart
potential, $V(x)=V_{0}[1+\tanh(x/a)]/2$ where $a$ depends on the extent of alloying. The reflective probability for this
potential can be found in the literature\cite{eckart} as:
\begin{align}
\big|r_{k_x,k^\prime_x}\big|^{2}=\frac{\cosh [2\pi
a(k_{x}-k^\prime_{x})]-1}{\cosh [2\pi a(k_{x}+k^\prime_{x})]-1}\label{d1}
\end{align}
where $k_{x}$ and $k^\prime_{x}$ are the longitudinal momenta in either
side of the interface.
Comparing with the sharp barrier, this modulation always diminishes the reflective probability. When $a$ is much
larger than the Fermi wavelength, Eq.(\ref{d1}) is suppressed exponentially as $\exp\big[-8\pi k_x^\prime a\big]$. 
According to Eq.(\ref{eq:interlayerE}), the magnetic coupling strength is proportional to the reflective probability, and so is expected to be reduced drastically by the alloying as concluded by the {\it ab initio} calculations\cite{alloy}.

If the interdiffusive process is weak, namely, $ k_{x}a\ll 1$, the alloying only introduces a small correction to the interlayer coupling strength:
\begin{align}
\triangle E^{a} \approx  \triangle E e^{-\frac{4}{3}(\pi a
)^{2}k_{F}k^\prime_{F}}
\end{align}
This means that the relative enhancement due to the correlations is the same as before considering the alloying.

\end{appendix}

\end{document}